\documentclass[english,twocolumn, superscriptaddress, amsmath]{revtex4}

\usepackage[colorlinks=true,linkcolor=blue,citecolor=blue]{hyperref}
\usepackage{times}
\usepackage{amsmath}
\usepackage{amssymb}
\usepackage{amsthm}
\usepackage{amsfonts}
\usepackage{enumerate}
\usepackage{latexsym}
\usepackage{ifpdf}
\usepackage{soul}
\usepackage{graphicx}
\usepackage{makeidx}
\usepackage{color}

\usepackage{graphicx}
\usepackage{dcolumn}
\usepackage{bm}

\usepackage{babel}

\begin{document}

\title{Effect of spin-orbit interaction  on the vortex dynamics in LaAlO$_3$/SrTiO$_3$ interfaces near the superconducting transition}

\author{Gopi Nath Daptary}
\thanks{Current address: Department of Physics, Bar Ilan University, Ramat Gan  5290002, Israel}
\affiliation{Department of Physics, Indian Institute of Science, Bangalore 560012, India}

\author{Hemanta Kumar Kundu}
\affiliation{Department of Physics, Indian Institute of Science, Bangalore 560012, India}
\author{Pramod Kumar}
\thanks{Current address: Department of Physics, St. John's College, Agra, Uttar Pradesh 282 002, India}
\affiliation{National Physical Laboratory, New Delhi 110012, India}
\author{Anjana Dogra}
\affiliation{National Physical Laboratory, New Delhi 110012, India}
\author{Narayan Mohanta}
\affiliation{Material Science and Technology Division, Oak Ridge National Laboratory, Oak Ridge, TN 37831, USA}
\author{A. Taraphder}
\affiliation{Department of Physics and Centre for Theoretical Studies, Indian Institute of Technology Kharagpur, W.B. 721302, India}
\author{Aveek Bid}
\email{aveek@iisc.ac.in}
\affiliation{Department of Physics, Indian Institute of Science, Bangalore 560012, India}

\begin{abstract}

Controlling spin-orbit interaction and its effect on superconductivity has been a long-standing problem in two-dimensional inversion symmetry broken superconductors. An open challenge is to understand the role of various energy scales in shaping the complex phase diagram in these systems.  From a combined experimental and theoretical study of resistance fluctuations and its higher order statistics, we propose a phase diagram for the superconducting phase in the magnetic-field--spin orbit interaction energy plane for the quasi-two dimensional electron gas at the interface of LaAlO$_3$/SrTiO$_3$ heterostructures. The relative variance of resistance fluctuations  increases by few orders of magnitude below the spin-orbit field B$_{SO}$ and a non-Gaussian component to the fluctuations arises for fields  below the upper critical field B$_{C2}$. Theoretical calculations show that the non-Gaussian noise predominantly arises due to percolative nature of the superconducting transition. We quantify the strength and the relative importance of the spin-orbit interaction energy, Zeeman energy and the pairing potential.  Our work highlights the important role played by the interplay between these energy scales in framing the  fascinating phases seen in two-dimensional inversion-symmetry-broken superconductors.

\end{abstract}

\maketitle
\section{Introduction}
Unconventional superconductivity is of great interest both from theoretical as well as experimental points of view \cite{sigrist1991phenomenological, fulde1964superconductivity}. Anderson's theorem~\cite{PhysRevB.30.4000} states that in the presence of both time-reversal and inversion symmetries one gets even-parity spin-singlet pairing in superconductors. The absence of either one of these symmetries - either through Zeeman effect (loss of time-reversal symmetry) or spin-orbit interaction (loss of inversion symmetry)  leads to the lifting of spin degeneracy favoring the formation of odd-parity spin-triplet cooper pairs~\cite{sigrist2009introduction}.  The effect of broken time-reversal symmetry on parity of cooper pairs is pretty well studied. There are several examples of superconductors in nature where the presence of magnetism leads to the appearance of non-trivial pairing - well known examples being heavy Fermion systems (e.g. CeIn$_3$ \cite{fukazawa2003theory},  CeCoIn$_5$ \cite{petrovic2001heavy} and UGe$_{2}$ \cite{huxley2001uge}),  iron-pnictides \cite{yin2009scanning}, certain organic superconductors. On the other hand, known examples of naturally occurring odd-parity pairing induced by spin-orbit interaction (SOI)  are much rarer - the obvious exceptions being non-centrosymmetric superconductors like CePt$_3$Si \cite{samokhin2004cept}, CeIrSi$_3$ \cite{tada2010spin} and CeRhSi$_3$ \cite{tada2010spin,kimura2007extremely}. Under certain conditions, odd-parity pairing can be induced in two-dimensional superconductors in the presence of SOI \cite{PhysRevLett.87.037004,michaeli2012superconducting}.

The quasi-two-dimensional  electron gas (q-2DEG) formed at the interface between (001) oriented SrTiO$_3$ and LaAlO$_3$ (hereafter referred as LaAlO$_3$/SrTiO$_3$) is one such system. Two factors lead to the appearance of a large Rashba SOI in this system: (a) breaking of parity symmetry at the interface, and (b) a large electric field perpendicular to the interface, primarily due to polar catastrophe (and to a lesser extent due to applied back-gate voltage). It is interesting to note that Rashba SOI has two notable consequences: (a) it induces charge inhomogeneity at the interface at sub-micron length scales~\cite{caprara2012intrinsic}, and (b) it induces an in-plane field perpendicular to the $k$-vector of the charge carriers \cite{zhong2013theory}. Both these factors are  expected to have a significant influence on superconductivity. Another  advantage of this q-2DEG over conventional non-centrosymmetric bulk superconductors is that both superconducting $T_C$ and SOI strength are gate-voltage tunable~\cite{shalom2010tuning, PhysRevLett.104.126803, caviglia2008electric}. A variety of exotic phenomena have been theoretically predicted to exist as a consequence of the SOI including  Fulde-Ferrell-Larkin-Ovchinikov-type (FFLO) superconductivity coexisting with ferromagnetism ~\cite{michaeli2012superconducting},  exotic superconducting pairing states which are an admixture of spin-singlet and spin-triplet components~\cite{PhysRevB.92.174531,0034-4885-80-3-036501} and emergent Majorana quasiparticles \cite{mohanta2014topological}.

In this paper we present detailed experimental studies of the effect of SOI on the magnetotransport and spin fluctuations in high-quality  LaAlO$_3$/SrTiO$_3$ heterostructures at temperatures much below the superconducting $T_C$. Study of second- as well as higher-order moments of  fluctuations of dynamical variables is a well established tool to probe the presence of long-range correlations in systems undergoing phase transitions~\cite{weissman1993spin,PhysRevLett.100.180601, koushik2013correlated, samanta2012non, daptary2014probing, daptary2016correlated, daptary2018effect}. From magnetotransport measurements we identify the relevant field scales: upper critical field B$_{c2}$ and spin-orbit field B$_{SO}$, which are gate voltage tunable. We observe that close to these field scales, resistance fluctuations and their higher order statistics develop strikingly non-trivial features. Both from experimental and theoretical data, we find that the interplay between spin-orbit interaction, pairing energy and Zeeman energy creates a fascinating phase diagram  very distinct from that usually found for conventional two-dimensional (2D) superconductors.\\

\section{EXPERIMENTAL DETAILS}
Our measurements were performed on samples with 10 unit cells of LaAlO$_{3}$ grown by pulsed laser deposition (PLD) on TiO$_{2}$ terminated (001) SrTiO$_{3}$ single crystal substrates. As received SrTiO$_{3}$ substrates were pre-treated with standard buffer hydrofluoric (NH$_4$F - HF) HF solution~\cite{kawasaki1994atomic} in order to achieve uniform TiO$_{2}$ termination. The TiO$_2$ termination of the substrate realized with the buffer HF solution etching was confirmed from atomic force microscopy measurements. Prior to deposition the treated substrates were annealed for an hour at 830$^\circ$C in oxygen partial pressure of 7.4 x 10$^{-2}$ mbar. The purpose of pre-annealing of substrates in oxygen atmosphere at 830$^\circ$C was to remove any moisture and organic contaminants from the surface and also to reconstruct the surface so that pure TiO$_2$ termination is realized. Further, 10 unit cells LaAlO$_{3}$ were deposited at 800$^\circ$C at an oxygen partial pressure of 1x 10$^{-4}$ mbar.  Growth with the precision of single unit cell was monitored by the oscillations count using in-situ RHEED gun. The epitaxial nature of the films was confirmed by HRXRD performed on a 20 unit-cell LaAlO$_3$ film grown under identical conditions on TiO$_2$ terminated SrTiO$_3$ which allowed us to measure the c-axis lattice parameter of LaAlO$_3$. The thickness of one unit cell from these measurements came out to be 3.75~\AA~\cite{kumar2015enhanced}. Ohmic electrical contacts were achieved by  ultrasonically bonding Au wires (25~$\mu$m diameter) at the four corners of the device in a van der Pauw geometry. This technique is known to breakdown the 10~u.c. of LaAlO$_3$ and provide ohmic contact with the underlying electron gas \cite{caviglia2008electric, joshua2012universal,han2014two,shalom2010tuning,daptary2016correlated, PhysRevB.95.174502, daptary2018effect}. All electrical measurements were performed in a cryogen-free dilution refrigerator over the temperature range 20--250~mK and magnetic field range 0--16~T. The relative angle between the magnetic field $B$ and the q-2DEG could be changed by rotating the sample \textit{in-situ} the dilution refrigerator and measurements were done with $B$ applied both parallel ($B_\parallel$) and perpendicular ($B_\perp$) to the interface. The charge carrier density at the interface was controlled using a back gate voltage $V_g$ with the SrTiO$_3$ acting as the dielectric material. Measurements were performed over the range $-200$~V$<V_g<$200~V. The interface was found to be superconducting for all values of $V_g>-10$~V.
\section{RESULTS AND DISCUSSION}
\begin{figure}[t]
	\begin{center}
		\includegraphics[width= 0.48\textwidth]{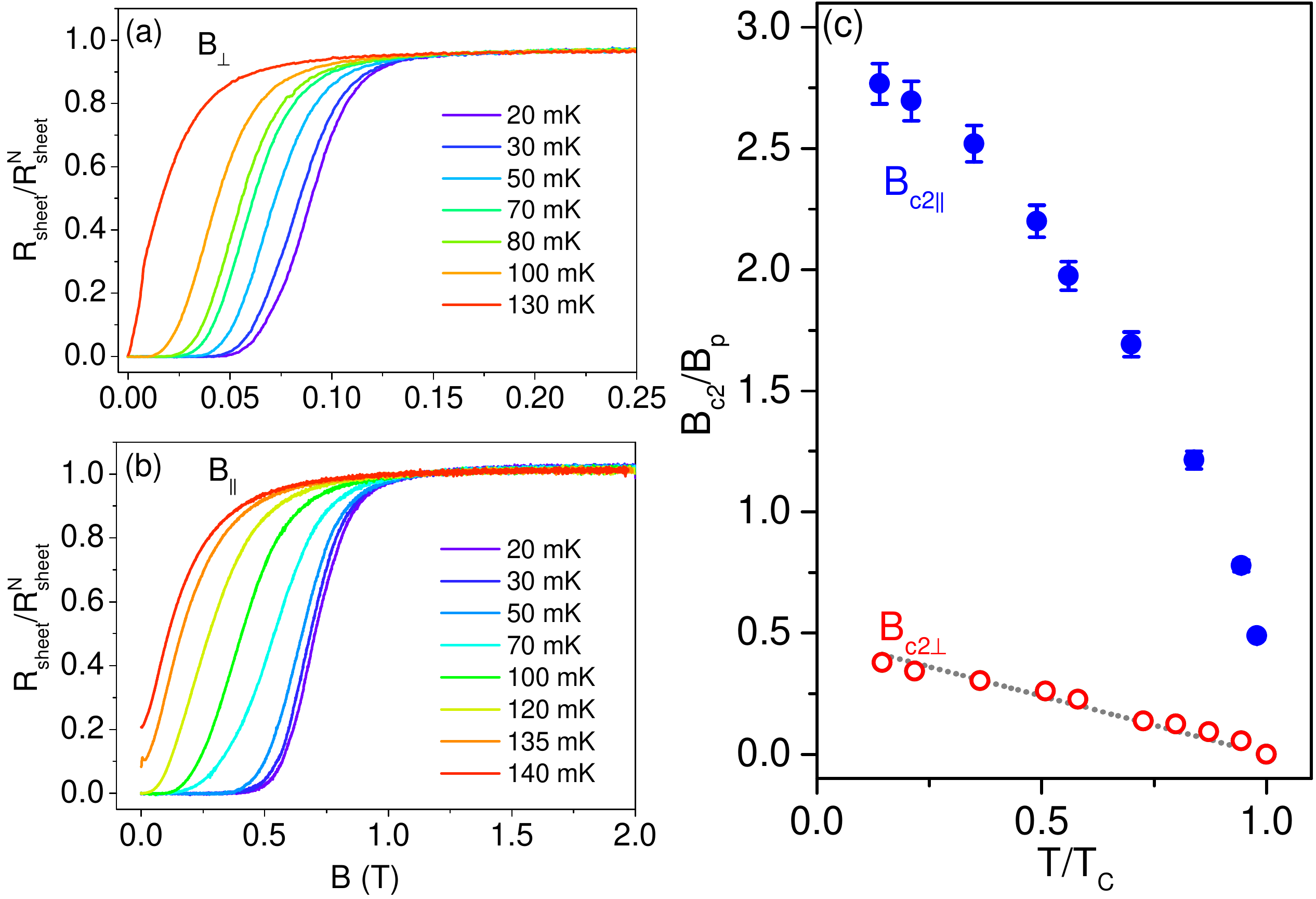}
		\small{\caption{Normalized sheet resistance versus temperature $T$ as a function of (a) perpendicular field, $B_\perp$ and (b) parallel magnetic field, $B_\parallel$. (c) Upper critical field $B_{c2}$ normalized by the Pauli paramagnetic field $B_p$ as a function of reduced temperature $T/T_C$ for fields applied parallel to the interface (blue filled circles) and perpendicular to the interface (red open circles). The gray dotted lines are fits to Eqn.~\ref{Eqn:GL1}. 
				The measurements were performed at $V_g=200$~V. 
				\label{fig:figure1}}}
	\end{center}
\end{figure}

We start with the results of magnetoresistance measurements at $V_g = 200$~V. The superconducting transition temperature $T_C$ (defined as the temperature where the zero field resistance became 40\% of its normal state value) was measured to be about 140~mK.  Figure~\ref{fig:figure1}(a) presents the normalized magnetoresistance $R_{sheet}/R_{sheet}^N$ as a function of perpendicular field $B_\perp$ at different temperatures for $V_g=200$ V. Here $R_{sheet}^N$ is the zero-field normal-state sheet resistance measured at $T=300$~mK. Fields of the order of 10~mT is enough to destroy the dissipationless superconducting state. The corresponding plots for $B_\parallel$ are shown in Fig.~\ref{fig:figure1}(b). As expected, given the quasi-2D nature of the system, the fields required in this case were at least two-orders of magnitude higher. 

In Fig. \ref{fig:figure1}(c) we plot the upper-critical field $B_{c2}$ (defined as the field at which the $R_{sheet}(B)$ drops to 40~\% of $R_{sheet}^N$) versus $T$ for both $B_\perp$  and $B_\parallel$. The values of $B_{c2}$ have been normalized by the BCS paramagnetic Pauli limit $B_p$, defined as  $\sqrt{2}g\mu_BB_p = 3.5k_BT_C$ \cite{chandrasekhar1962note,clogston1962upper}; $g$ being the gyromagnetic ratio, $k_B$  the Boltzmann constant  and $\mu_B$ the Bohr magneton. The dependence of $B_{c2}$ on the temperature $T$ for the out-of-plane  is fitted well by the phenomenological 2D Ginzburg-Landau model \cite{tinkham1963effect}
\begin{align}
	B_{c2\perp}=\frac{\Phi_0}{2\pi \xi_{GL}(0)^2}(1-T/T_c)
	\label{Eqn:GL1}
\end{align}

%
where $\xi_{GL}(0)$ is the in-plane GL coherence length at $T=0$ K, $\Phi_0=h/2e$ is the flux quantum. The value of $\xi_{GL}(0)$ extracted from the fit is 55~nm which matches well with previous reports \cite{reyren2009anisotropy,shalom2010tuning}. From Fig. \ref{fig:figure1}(c) we observe that $B_{c2\parallel}$ far exceeds the Clogston-Chandrashekhar limit which, in the weak coupling approximation, is expected to limit the value of the parallel upper critical field to $B_{c2\parallel} \leq B_p$. This large enhancement of  $B_{c2\parallel}$ has been reported previously in (001) LaAlO$_3$/SrTiO$_3$ hetero-interfaces~\cite{shalom2010tuning}  and has been postulated to arise from the presence of strong Rashba SOI which weakens spin paramagnetism by mixing the quasiparticle  spin states~\cite{PhysRevB.12.877, PhysRevB.25.171}. Other possible mechanisms like anisotropic pairing mechanism, strong-coupling superconductivity or other exotic many-body effects have been considered and ruled out by previous workers (see for example~\cite{PhysRevLett.119.237002, shalom2010tuning}). 

For the case of strong SOI, $B_{c2\parallel}$ is related to the spin-orbit scattering time through~\cite{PhysRevB.12.877}:
\begin{eqnarray}
	\tau_{SO}=0.362\frac{\hbar}{k_BT_c}\Big(\frac{B_P}{B_{c2\parallel}(0)}\Big)^2
	\label{Eqn:comparison}
\end{eqnarray}
Using this relation yields $\tau_{SO}=4\times 10^{-13}$~s for $V_g$=170~V. 

\begin{figure}[t]
	\begin{center}
		\includegraphics[width= 0.48\textwidth]{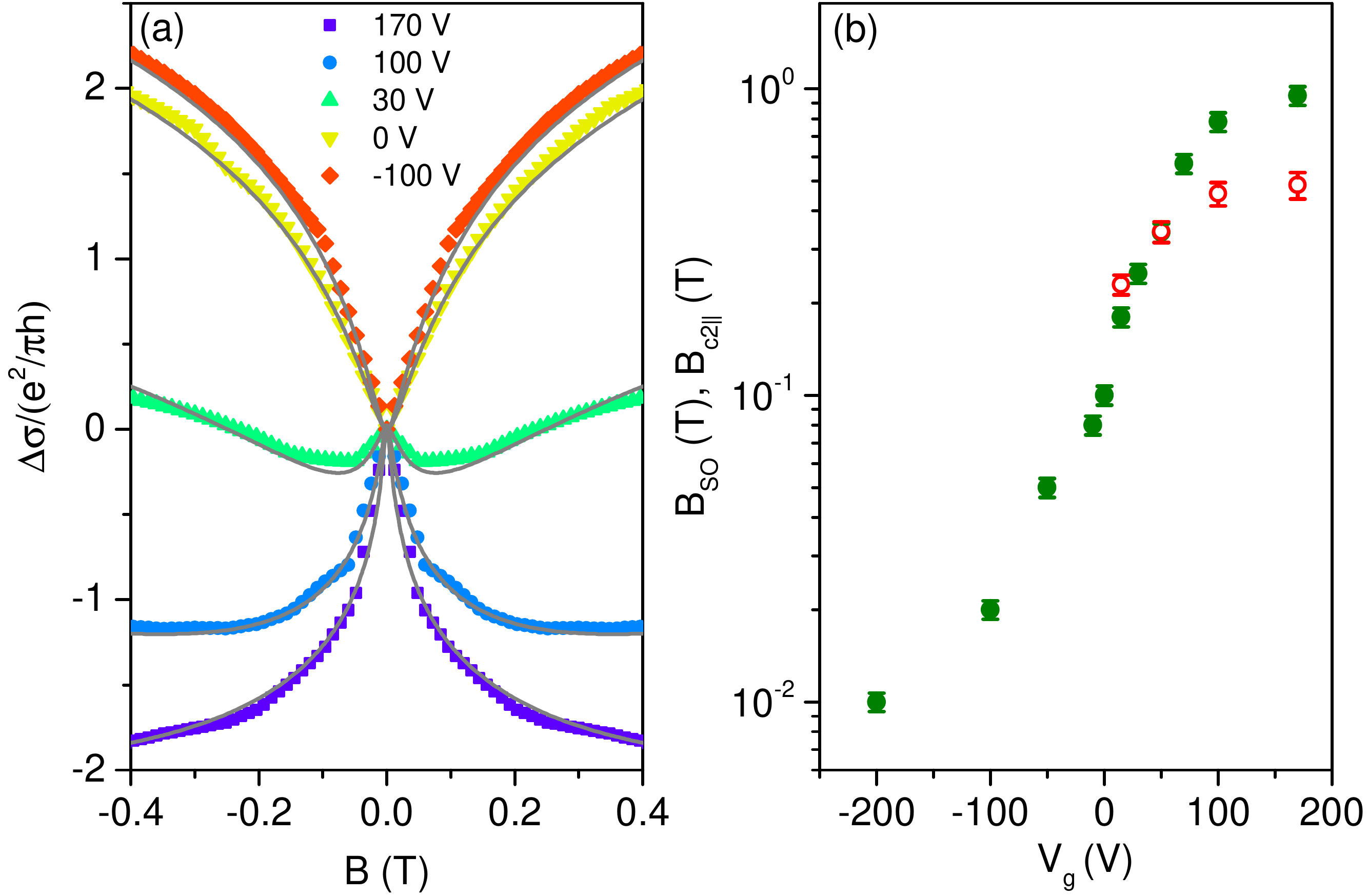}
		\small{\caption{(a) Magnetoconductance as a function of $B_\perp$ at different values of $V_g$. The scatter points are the measured data points while the solid lines are fits to the  Eqn.~\ref{Eqn:WAL}. (b) Plot of $B_{SO}$ (olive filled circles) and  $B_{c2\parallel}$ (red filled circles) versus $V_g$.  The measurements were performed at 245~mK. \label{fig:figure2}}}
	\end{center}
\end{figure}

The  SOI strength can also be  extracted from the measured low-field magnetoconductance at $T>T_C$. In a two dimensional system with in-plane SOI, in the presence of a perpendicular magnetic field $B_\perp$, the correction to conductance $\Delta \sigma$ takes the Maekawa-Fukuyama  form~\cite{maekawa1981magnetoresistance}:
\begin{align}
	\Delta\sigma(B)= &\frac{e^2}{\pi h}\Big[\Psi\Big(\frac{B_\perp}{B_i+B_{SO}}\Big) \notag \\
	&+\frac{1}{2\sqrt {1-\gamma^2}} \Psi\Big(\frac{B_\perp}{B_i+B_{SO}(1+\sqrt{1-\gamma^2}}\Big) \notag \\
	&-\frac{1}{2\sqrt {1-\gamma^2}} \Psi\Big(\frac{B_\perp}{B_i+B_{SO}(1-\sqrt{1-\gamma^2}}\Big)\Big].
	\label{Eqn:WAL}
\end{align}
Here $\Psi(x)$ = ln(x)+$\psi(0.5+\frac{1}{x})$, where $\psi$ is the digamma function. $B_i = \hbar/(4eD\tau_i)$ and $B_{SO} = \hbar/(4eD\tau_{SO})$ are inelastic and spin-orbit fields respectively ($\tau_i$ and $\tau_{SO}$ are respectively the inelastic and spin-orbit scattering times), $D$ is the diffusion constant and $\gamma$ is the Zeeman correction $\gamma=g\mu_B B/4eDB_{SO}$ ($g$ and $\mu_B$ are the electron $g$ factor and Bohr magnetron respectively).

The low-field magnetoconductance at $T=245$~mK is plotted in Fig.~\ref{fig:figure2}(a).   From the fits to these curves we extract the $\tau_{SO}$ and  $B_{SO}$. The value of $\tau_{SO}$ extracted from the fits to the magnetoresistance measured at $V_g$=170~V is $1.6\times10^{-13}$~s which matches closely with the value extracted using Eqn.~\ref{Eqn:comparison}. The value of $\tau_{SO}$, $\tau_{i}$ and $\tau_{elas}$ (elastic scattering time) are shown in Fig. \ref{fig:appendix1} in Appendix. As shown in Fig.~\ref{fig:figure2}(b), the value of $B_{SO}$ increases by almost two orders of magnitude as $V_g$ is swept from -200~V to 200~V. At low  $V_g$, $B_{SO}$ and $B_{c2\parallel}$ are comparable (Fig.~\ref{fig:figure2}(b)). With increasing $V_g$, $B_{SO}$ increases rapidly and becomes significantly larger than $B_{c2\parallel}$.  

\begin{figure}[t]
	\begin{center}
		\includegraphics[width=0.48\textwidth]{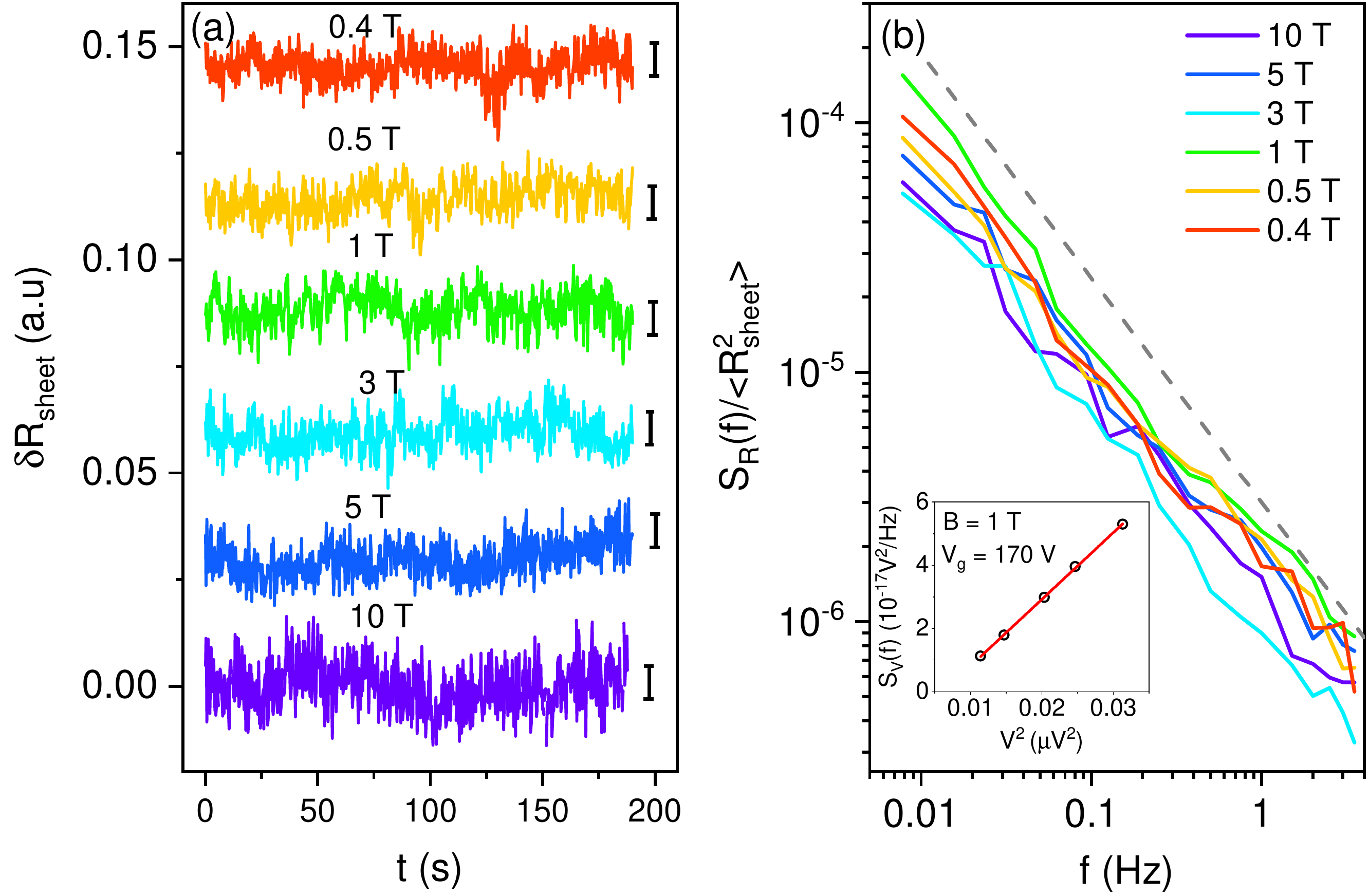}
		\small{\caption{(a) Time series of resistance fluctuations at a few representative values of $B_\parallel$. The measurement was performed at $T=20$ mK and $V_g=200$~V. (b) PSD of resistance fluctuations corresponding to the time-series shown in (a). Inset: PSD is plotted as a function of $V^2$ at $B=1$ Tesla and $V_g=170$ V - the linear dependence of $S_V(V)$ on $V^2$ establishes that noise originates from resistance fluctuations of the sample. \label{fig:figure3}}}
	\end{center}
\end{figure}

To probe the effect of spin-orbit interactions on charge carrier dynamics in the superconducting state, we studied resistance fluctuations for different  magnetic fields at $T=20$~mK ($T/T_C\approx 0.1$). The measurements were performed using a standard four-probe ac measurement technique (For details see Ref.~\cite{ghosh2004set}). Briefly, at each value of $V_g$ and $B$, the device is biased by a small ac current and the time series of resistance fluctuations $\delta R_{sheet}(t)$ is measured for 30 min using a dual-phase digital lock-in amplifier. The output of the lock-in amplifier is recorded by a fast data acquisition (DAQ) card. After extensive digital filtering of $\delta R_{sheet}(t)$ to remove line frequency and aliasing-effects, the power spectral density (PSD) of resistance fluctuations $S_R(f)$ was calculated using the method of Welch Periodogram.  The time-series of resistance fluctuations for a few representative values of $B_\parallel$, measured at $T$=20~mK and $V_g$=200~V, are plotted in Fig.~\ref{fig:figure3}(a). The corresponding PSD are shown in Fig.~\ref{fig:figure3}(b). For all values of $V_g$ and $B$, the dependence of $S_R(f )$ on the frequency $f$ was  found to be of the form $S_R(f) \propto 1/f^\alpha$ with $\alpha \sim 0.9-1$. $S_V(f)$ was always found to depend quadratically on the voltage $V$ developed across the channel [see inset of Fig~\ref{fig:figure3}(b)] establishing that the measured noise  originated from resistance fluctuations of the sample.

\begin{figure}[t]
	\begin{center}
		\includegraphics[width=0.48\textwidth]{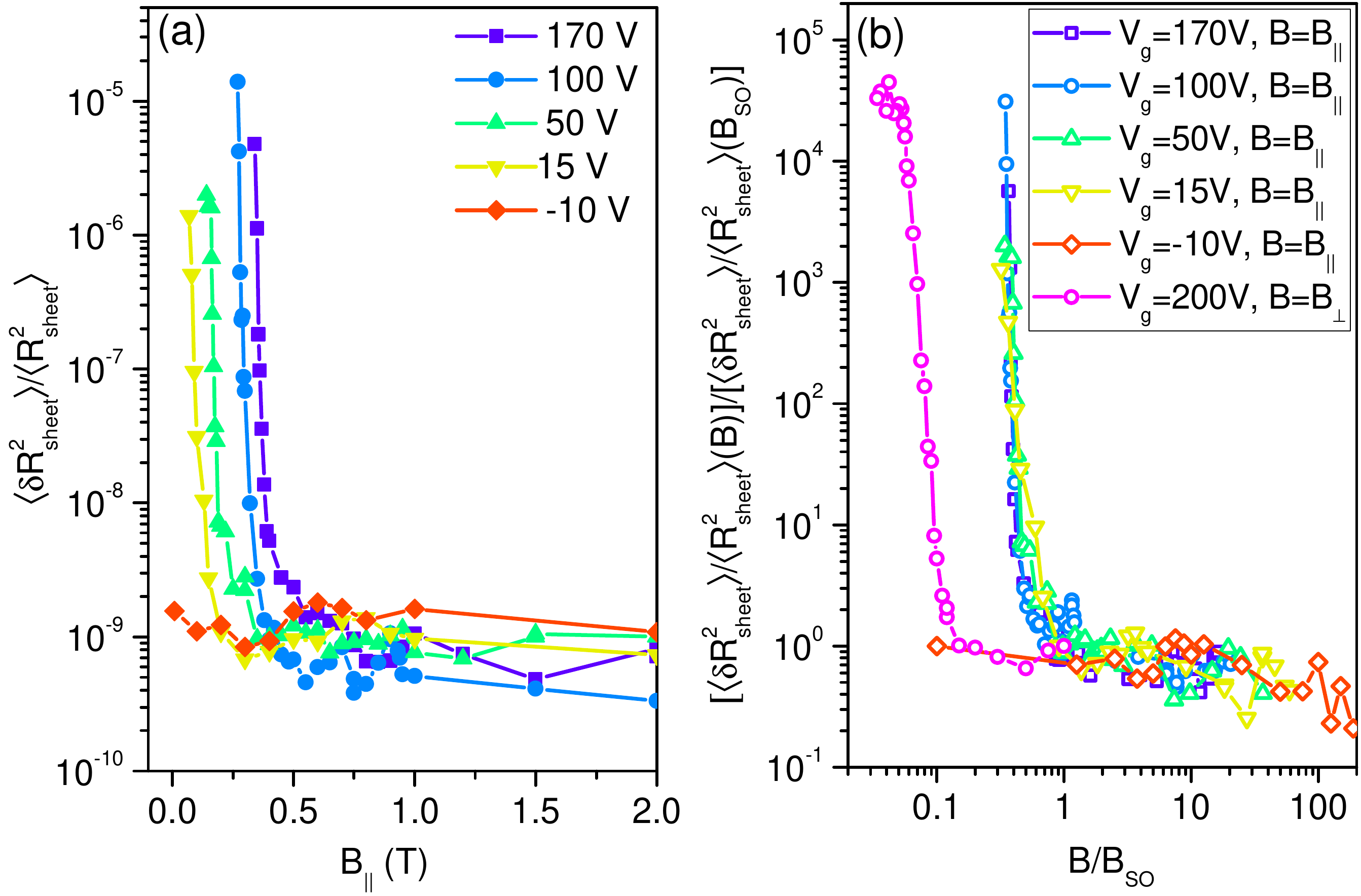}
		\small{\caption{(a) Plots of relative variance of resistance fluctuations $\mathcal{R}(B)$ versus $B_\parallel$, at different values of $V_g$. The measurements were performed at 20~mK. (b) Scaling plot of noise $\mathcal{R}(B)/\mathcal{R}(B_{SO})$ versus $B_\parallel/B_{SO}$ for V$_g$= 170 V, 100 V, 50 V, 15 V and -10 V respectively. Also, plotted is  $\mathcal{R}(B)/\mathcal{R}(B_{SO})$ versus $B_\perp/B_{SO}$ for field applied perpendicular to the interface at $V_g=200$ V (magenta open circles).  \label{fig:figure4}}}
	\end{center}
\end{figure}

The PSD of resistance fluctuations was integrated over the measurement bandwidth (7~mHz-4~Hz) to obtain the relative variance of resistance fluctuations $\mathcal{R}$:
\begin{align}
	\mathcal{R} \equiv \frac{\langle \delta R_{sheet}^2 \rangle}{\langle R_{sheet}^2\rangle}=\frac{1}{\langle R_{sheet}^2\rangle}\int S_R(f)df 
	\label{PSD}
\end{align}
In Fig. \ref{fig:figure4}(a) we show the plots of relative variance of resistance fluctuations $\mathcal{R}$ as a function of $B_\parallel$ at a few representative values of $V_g$ at $T=20$~mK. At high $B_\parallel$, the noise has a very shallow dependence on the field. Below a certain characteristic field, which is specific to $V_g$, the noise increases rapidly with decreasing $B$. Normally, one would expect this characteristic field to be the upper critical field, above which superconducting fluctuations are suppressed. However, a closer inspection of the data reveals that the characteristic field scale in this case is the spin-orbit field $B_{SO}$. As the field decreases below $B_{SO}$, the noise increases rapidly -- growing by over four orders of magnitude in the narrow magnetic field range $B_{c2\parallel}<B_\parallel<B_{SO}$.
In Fig.~\ref{fig:figure4}(b) we show a scaling plot of the noise $\mathcal{R}(B)/\mathcal{R}(B_{SO})$ as a function of $B/B_{SO}$. The data for all $V_g>-10$~V collapse onto  a single curve showing that indeed $B_{SO}$ is the relevant scale governing the $B_\parallel$ dependence of the resistance fluctuations in a superconductor with strong SOI. 

\begin{figure}[t!]
	\begin{center}
		\includegraphics[width=0.48\textwidth]{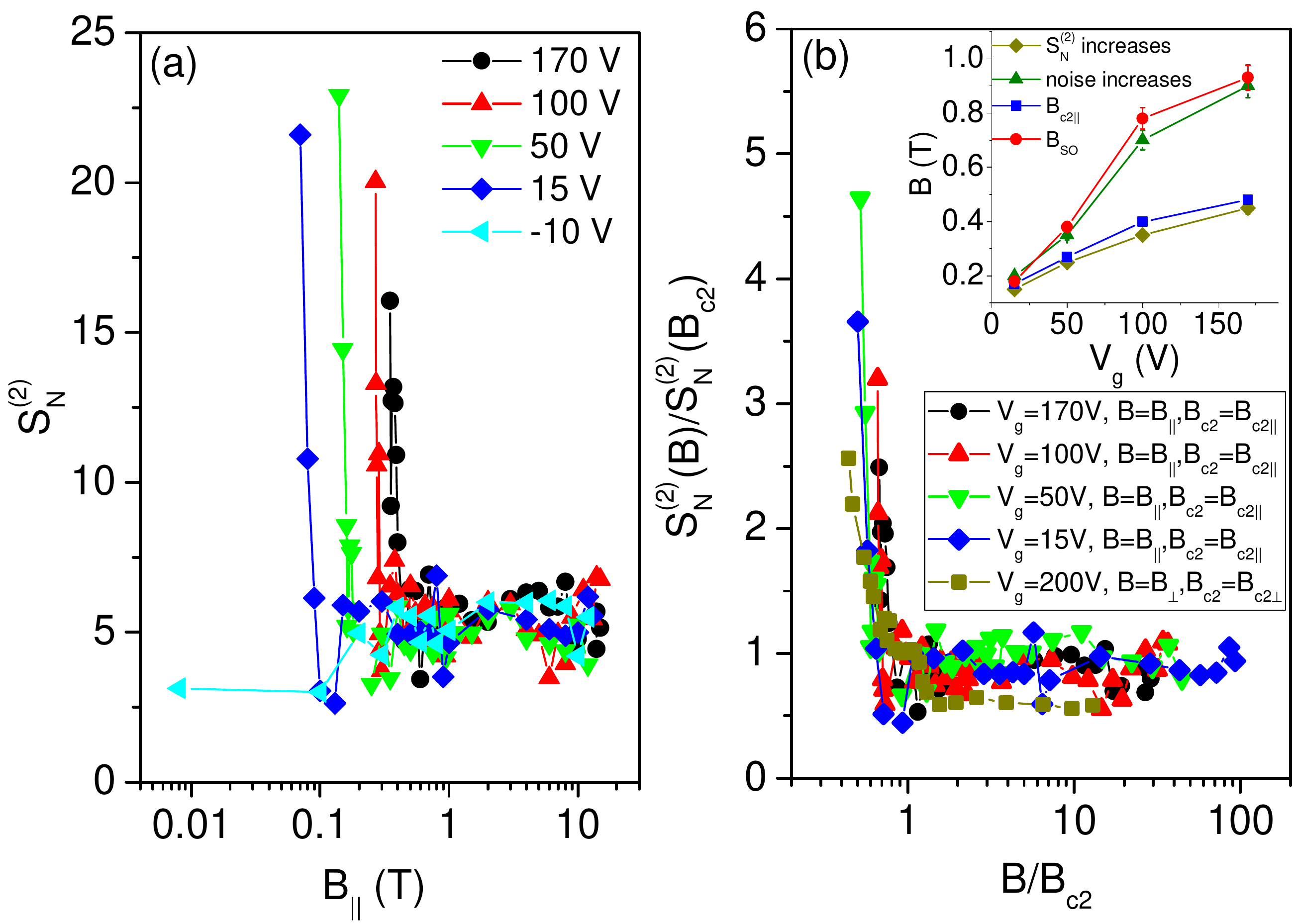}
		\small{\caption{ (a) Plot of $S_N^{(2)}$ as a function of magnetic field applied parallel to the interface at different values of gate voltages. The measurement was performed at 20 mK. (b) Scaling plot of $S_N^{(2)}(B_\parallel)/S_N^{(2)}(B_{c2\parallel})$ versus $B_\parallel/B_{c2\parallel}$ for $V_g$=170 V, 100 V, 50 V and 15 V respectively.  Also, plotted is  $S_N^{(2)}(B_\perp)/S_N^{(2)}(B_{c2\perp})$ versus $B_\perp/B_{c2\perp}$ for field applied perpendicular to the interface at $V_g=200$ V (dark yellow squares). \label{fig:figure5}}}
	\end{center}
\end{figure}

\begin{figure}[t!]
	\begin{center}
		\includegraphics[width=0.48\textwidth]{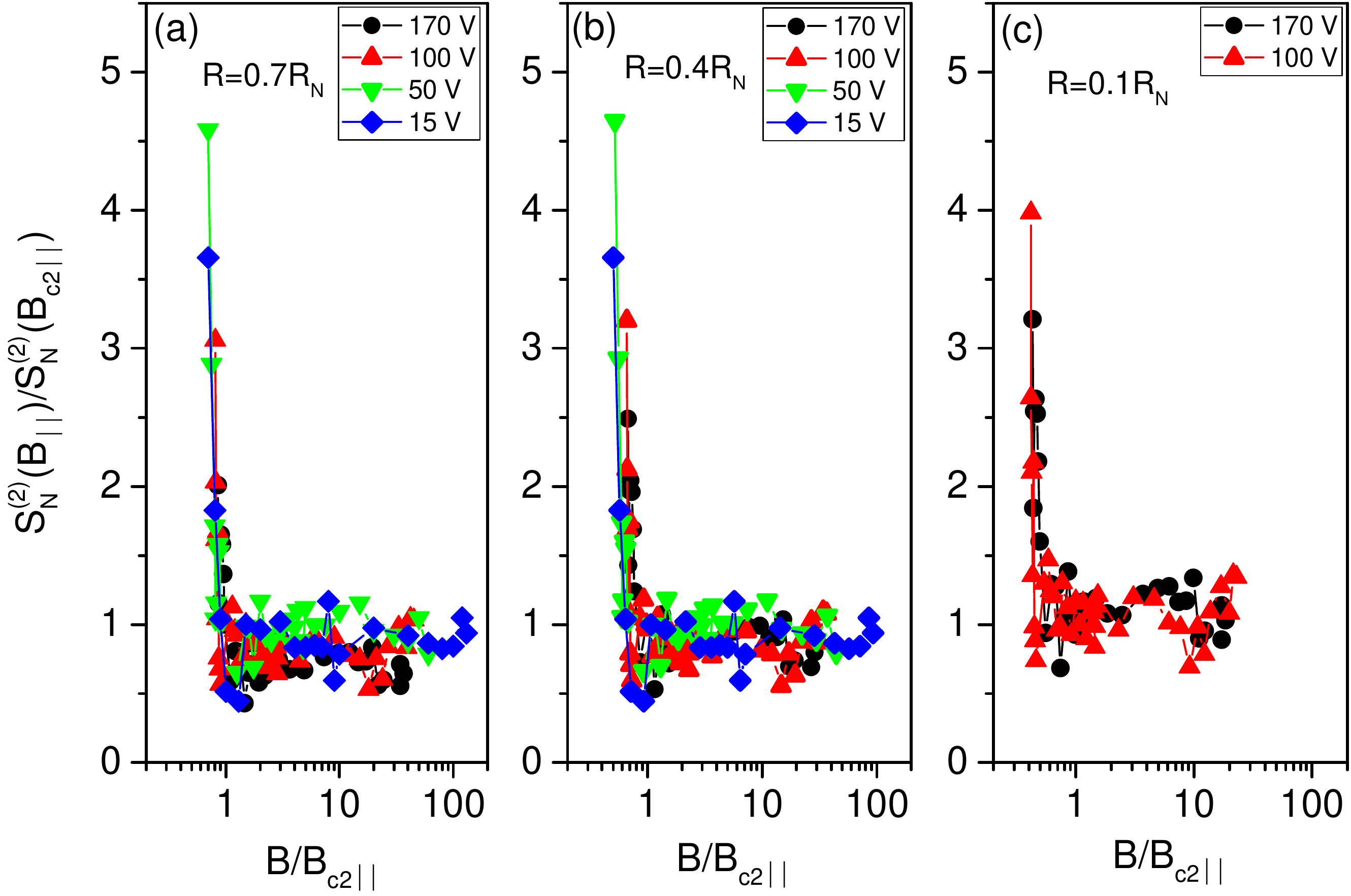}
		\small{\caption{ Scaling plot of $S_N^{(2)}(B)/S_N^{(2)}(B_{c2\parallel})$ versus $B_\parallel/B_{c2 \parallel}$. $B_{c2}$ are defined from the resistance transition at (a) $0.7R_{sheet}^N$, (b) $0.4R_{sheet}^N$ and (c) $0.1R_{sheet}^N$ resistance criterion.  \label{fig:appendix2}}}
	\end{center}
\end{figure}

To understand the origin of the measured resistance fluctuations, we studied their higher-order statistics. Such studies have been used extensively to detect the presence of long-range correlations in systems undergoing magnetic, spin-glass or superconducting transitions~\cite{weissman1993spin, PhysRevLett.100.180601, koushik2013correlated, samanta2012non, daptary2014probing, daptary2016correlated, daptary2018effect}. The Central Limit Theorem states that for uncorrelated random fluctuators, the fluctuation statistics is Gaussian. As the correlation length in the system begins to diverge  near a critical phase transition, the resultant time-dependent fluctuation statistics becomes strongly non-Gaussian~\cite{weissman1993spin, PhysRevLett.100.180601, koushik2013correlated, daptary2018effect}.  We computed the `second spectrum'  which is the four-point correlation function of the resistance fluctuations over a chosen frequency octave ($f_l, f_h$)~\cite{PhysRevB.31.2254, PhysRevB.53.9753}. It is mathematically defined as
\begin{equation}
	S_R^{f_1}(f_2)=\int_0^\infty \langle\delta R^2(t)\rangle\langle\delta R^2(t+\tau)\rangle \cos(2\pi f_2\tau)d\tau
	\label{SP}
\end{equation}
where $f_1$ is the center-frequency of the chosen octave and $f_2$ the spectral frequency. Physically, $S_R^{f_1}(f_2)$ represents `spectral wandering' of the PSD with time. To avoid corruption of the signal by the Gaussian background noise, the second spectrum was calculated over the frequency octave 93.75--187.5 mHz, where the sample noise is significantly higher than the background noise. A convenient way of representing the second spectrum is through its normalized form $S_N^{(2)}$ defined as
\begin{equation}
	S_N^{(2)}=\int_0^{f_h-f_l}S_R^{f_1}(f_2)df_2/[\int_{f_l}^{f_h}S_R(f)df]^2
	\label{NSP}
\end{equation}
For Gaussian fluctuations, $S_N^{(2)}$ = 3. The measured values of $S_N^{(2)}$  as a function of $B_\parallel$ is shown in Fig. \ref{fig:figure5}(a). We see that as the magnetic field is decreased below $B_{c2\parallel}$, $S_N^{(2)}$ starts increasing monotonically from its high field value which was close to 3. This can be appreciated better from Fig.~\ref{fig:figure5}(b) where we plot $S_N^{(2)}(B_\parallel)/S_N^{(2)}(B_{c2\parallel})$ as a function of $B_\parallel/B_{c2\parallel}$. The data for all $V_g$ collapse onto a single plot showing  that the relevant field scale for the second spectrum is $B_{c2\parallel}$. We note that scaling plot $S_N^{(2)}(B_\parallel)/S_N^{(2)}(B_{c2\parallel})$ remains unchanged if $B_{c2\parallel}$ are defined for other resistance criterion, e.g., $R_{sheet} = 0.7R_{sheet}^N$ and $R_{sheet} = 0.1R_{sheet}^N$ (see Fig. \ref{fig:appendix2} ). For $V_g= -10$ V, where the device is in resistive state over the entire magnetic field range, the relative variance of resistance fluctuations $\mathcal{R}$ is independent of  field (Fig.~\ref{fig:figure4}) and $S_N^{(2)}\simeq3$ (Fig.~\ref{fig:figure5}) showing that the fluctuations in normal state in LaAlO$_3$/SrTiO$_3$ are Gaussian. \\

To summarize our observations so far: (a) for $B_\parallel>B_{SO}$, the resistance fluctuations are almost independent of $B_\parallel$  and have a Gaussian distribution, (b) there is a significant range of field $B_{SO}>B_\parallel>B_{c2\parallel}$ where the resistance fluctuations  depend strongly on $B_\parallel$ while remaining Gaussian, and (c) for $B_\parallel < B_{c2\parallel}$, the resistance fluctuations are large, have a strong $B_\parallel$-dependence and have a non-Gaussian distribution. In the inset of Fig.~\ref{fig:figure5}(b) we summarize the data.  One can see that the magnetic field at which the second spectrum deviates from the Gaussian value (we call it $B_{NG}$) closely follows the upper critical field $B_{c2 \parallel}$ while the field at which the noise begins to shoot up (labeled $B_N$) tracks $B_{SO}$. At this point it is profitable to compare these observations with what is seen for $B_\perp$ for this q-2DEG  superconductor --  a representative data taken at $T$ = 20~mK and $V_g$ = 200~V is plotted in Fig.~\ref{fig:figure4}(b) (magenta open circles).  For $B_\perp>B_{c2\perp}$, the noise is field-independent, small in magnitude and Gaussian. For $B_\perp<B_{c2\perp}$,  the noise is non-Gaussian and diverges strongly as the field is reduced ( Fig. \ref{fig:figure5}(b) - dark yellow squares). Notably, in contrast to $B_\parallel$, the divergence of noise and appearance of non-Gaussian component are concurrent. This has been observed previously in other 2-dimensional superconductors and has been shown to arise due to long range correlations between the vortices near the transition \cite{PhysRevLett.76.2551,PhysRevLett.78.519, PhysRevB.57.6036, Wagenblast1998, daptary2016correlated, koushik2013correlated}. 

\begin{figure}[t!]
	\begin{center}
		\includegraphics[width=0.48\textwidth]{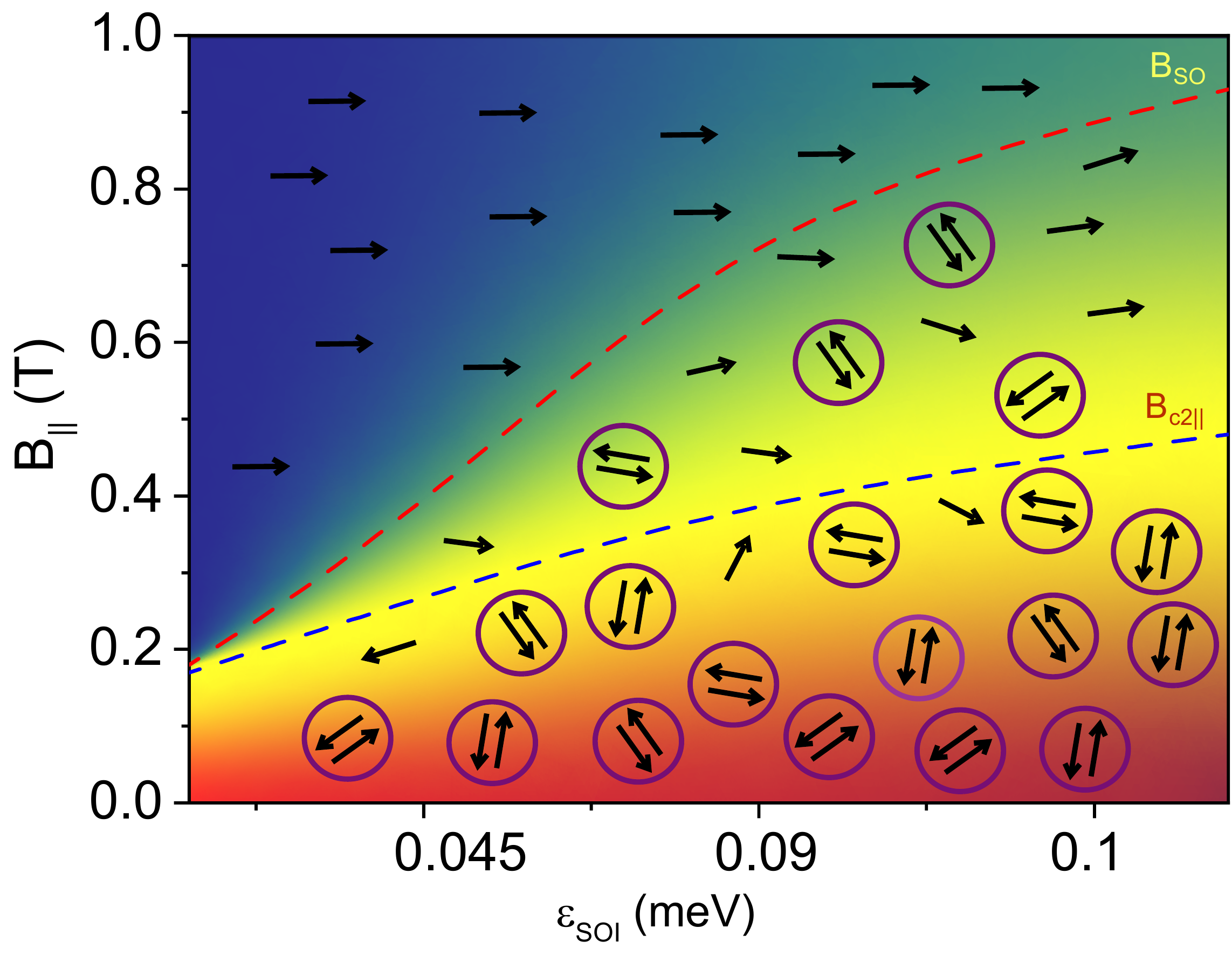}
		\small{\caption{ Schematic phase diagram showing the spin orientations at the  LaAlO$_3$/SrTiO$_3$ interface.  Upper critical field B$_{c2 \parallel}$ (blue line) and spin-orbit field B$_{SO}$ (red line) have been plotted in the SOI-energy $\epsilon_{SOI}$ and $B_\parallel$ plane. Arrows indicate the direction of spin of a single electron in the plane of the q-2DEG while circles represent the cooper pairs.  \label{fig:figure6}}} 
	\end{center}
\end{figure}

We now discuss the possible origin of the decoupling of $B_{NG}$ and $B_N$ in this system. As shown before~\cite{koushik2013correlated, daptary2016correlated}, it is correlations between vortices  that leads to non-Gaussian noise in 2D-superconductors. Thus, it is natural that $B_{c2 \parallel}$  (the field at which superconductivity is destroyed)  and $B_{NG}$ (the field at which non-Gaussian fluctuations vanish) coincide. The measured resistance fluctuations, however, persist beyond $B_{c2 \parallel}$ deep into the normal state, until $B_\parallel\sim B_{SO}$. Below we present a plausible scenario which explains this. Strong SOI present in this system ensures that the electronic spins are all in-plane. As the electronic transport is diffusive, the $k$-vector of the charge carriers take random values. Spin-momentum locking due to SOI causes these charge carriers to feel an effective in-plane $B_{SO}$ field perpendicular to the $k$-vector. The competition of this random $B_{SO}$ with $B_\parallel$  brings down the in-plane spin magnetic moment to $\sim( B_\parallel/B_{SO})\mu_B$~\cite{xi2016ising, youn2012role, sigrist2009introduction}. At large enough parallel magnetic fields, Zeeman energy ensures that all the spins are aligned along $B_\parallel$. As $B_\parallel$ is reduced to the order of $B_{SO}$ there begins to appear spins of opposite signs which can form Cooper pairs. With further reduction of $B_\parallel$, the superfluid density grows and for  $B_\parallel <B_{c2\parallel}$ global phase coherence sets in. Thus, in the field range $B_{SO}>B_\parallel>B_{c2\parallel}$ there will exist domains of superconducting clusters in a background of normal carriers. We propose that it is fluctuations of these superconducting clusters that give rise to the large Gaussian  noise over this field regime. We present a schematic phase diagram of the spin orientation in Fig.~\ref{fig:figure6} in the SOI-energy $\varepsilon_{SOI}$ and $B_\parallel$ plane. The values of $\varepsilon_{SOI}$ have been obtained from $\tau_{SO}$ extracted from the fits to the magnetoresistance data at different $V_g$ using Eqn.~\ref{Eqn:WAL}. This picture is in some sense analogous to what one gets in the zero-field limit - as the temperature is reduced sufficiently close to $T_C$, there appears percolating clusters with finite superfluid density in a resistive background which gives rise to large Gaussian resistance fluctuations. It has been predicted that FFLO state is favorable in the phases between $B_{c2}$ and $B_{SO}$ \cite{barzykin2002inhomogeneous,dimitrova2007theory}, which possibly, can have contributions to the resistance fluctuations. Without experimental data, we refrain commenting on it.

 The magnetic field-induced transition to the superconducting state is affected by non-magnetic disorder \cite{mohanta2014oxygen, mohanta2013phase} and the transition is assumed to be percolative in nature. To describe such a percolative phase transition induced by in-plane magnetic field $B$, a random resistor network (RRN) model was considered \cite{PhysRevLett.54.1718,PhysRevLett.54.2529}. In this model, we consider a square network of identical resistors of size $L \times L$, where $L$ is the number of grid points along $x$ or $y$ direction. In the ideal scenario, the resistor network is assumed to be connected by external conducting wires to a voltage source $V$, which causes a current $I$ to flow through the network, so that the macroscopic sheet-resistance is measured as $R_{sheet} = V/I$. In this model, we discretize the resistance and define the mean resistance at a grid point ($x_i,y_i$) by $R_i$, so that the macroscopic resistance is given by averaging over all grid points in the network \textit{viz.} $R_{sheeet}=(1/L^2)\sum_i R_i$.
\begin{figure}[t!]
	\begin{center}
		\includegraphics[width=0.48\textwidth]{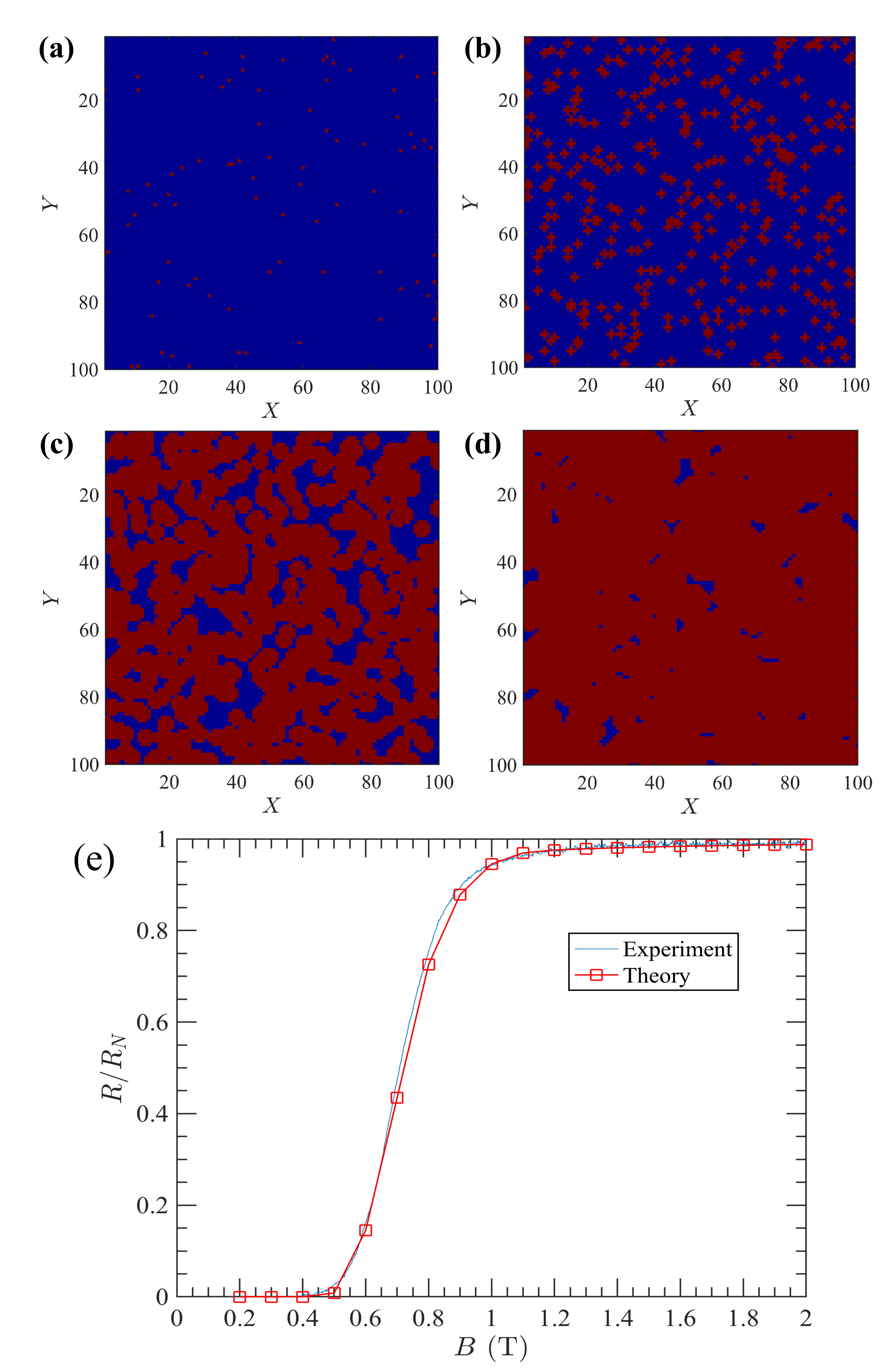}
		\small{\caption{ Colormap of the sheet-resistance $R_{sheet}$ in a $100\times100$ network  in the RRN model at different in-plane magnetic fields (a) $B_\parallel=1$ T, (b) $B_\parallel=1.3$ T, (c) $B_\parallel=1.6$ T and (d) $B_\parallel=1.9$ T, across the transition from superconducting state to the normal metallic state. Blue background denotes regions with resistance $R_{sheet}=0$, while red dots/patches denote regions with high resistance $R_{sheet}=R_{sheet}^N$ ($R_{sheet}=R_{sheet}^N$ being the resistance in the normal metallic state). In this plot, temperature $T=20$~mK and gate-voltage $V_g=170$~V. (e) The blue-line shows the variation of the normalized resistance $R_{sheet}/R_{sheet}^N$ (blue curve) with in-plane magnetic field $B_\parallel$ at  $T=20$~mK and  $V_g=170$~V, where $R_{sheet}^N$ is the resistance in the normal metallic state. The theoretical fit obtained using the RRN model is shown by the red open circles. \label{fig:Res_map}}}
	\end{center}
\end{figure}
\begin{figure}[t!]
	\begin{center}
		\includegraphics[width=0.48\textwidth]{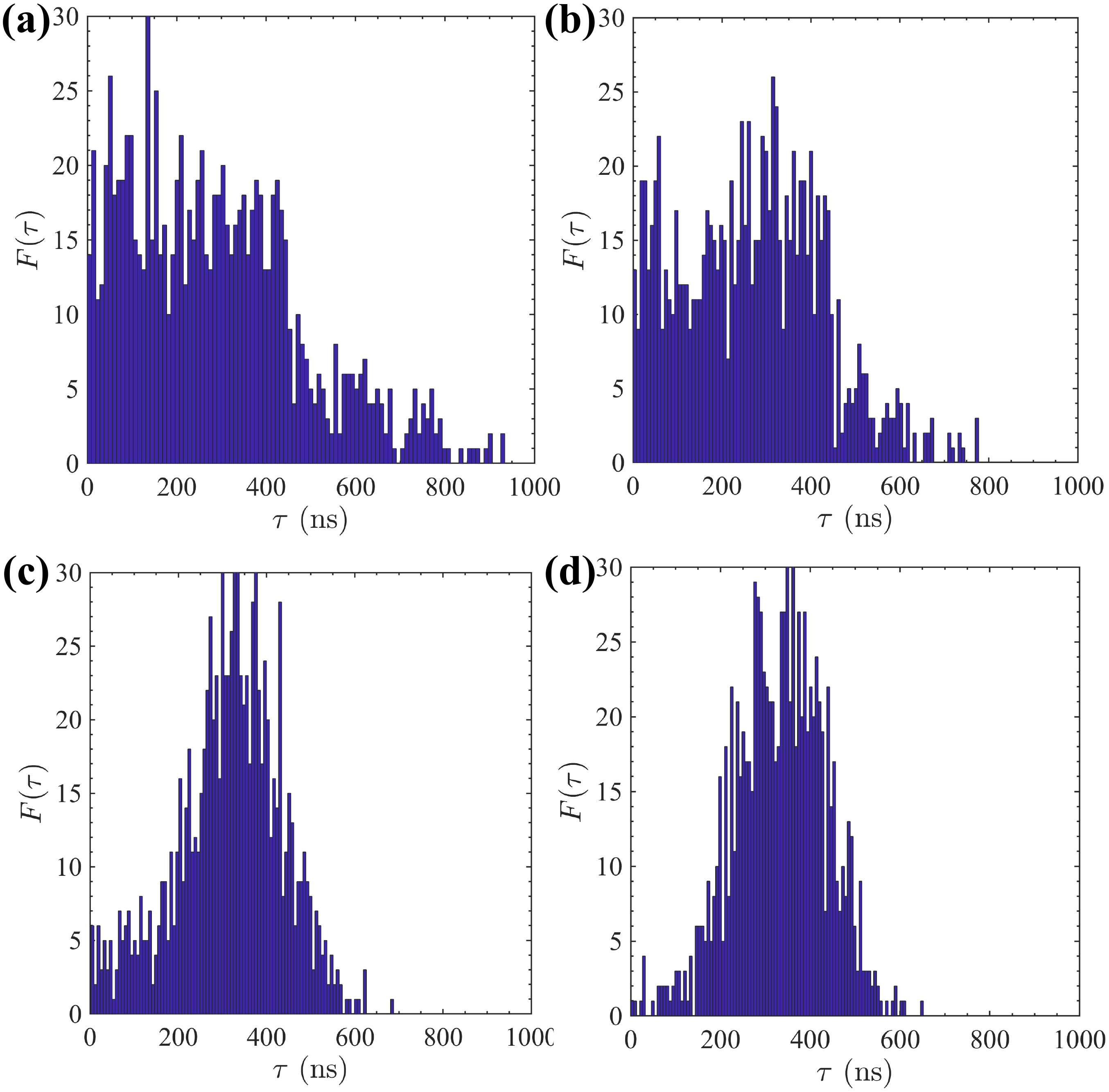}
		\small{\caption{Distribution of the relaxation time at different in-plane magnetic field $B_\parallel$ values (a) 1 T, (b) 1.3 T, (c) 1.6 T and (d) 1.9 T. The distribution changes from non-Gaussian type to Gaussian type as $B_\parallel$ is increased across the transition from superconducting state to normal-metallic state. In this plot,  $T=20$~mK and  $V_g=170$~V. \label{fig:tau_dist}}}
	\end{center}
\end{figure}

In the RRN model, we consider a $100\times100$ network in which circular resistive clusters appear in the superconducting phase when $B$ is increased, as shown in Fig. \ref{fig:Res_map}. The $B$-dependence of the number and diameter of the clusters are given, respectively, by $N_{cluster}=Int.(C_1(B-B_c))$ and $D_{cluster}=C_2B_r$, where $B_r=(B-B_c)/B_c$, $B_c$ is the critical field for the superconducting transition (at a given value of gate voltage $V_g$, we take $B_c$ as the highest available critical field $i.e.$ $B_c=B_{so}$), $C_1$ and $C_2$ are parameters which are determined by fitting $R$ with experimental data, the function $Int.()$ returns the integer value of the number inside the bracket. The value of the resistance inside the resistive clusters is large, here we assume $R_{sheet}=R_{sheet}^N$, the value in the normal metallic state at $B=2$~T. The normalized resistance at a field $B$ is given by $R/R_N=1/(1+\xi^2)$, where $\xi$ is the superconducting coherence length. We assume that in a disordered BCS superconductor with percolative superconducting transition, $\xi$ follows a field dependence which is similar to the temperature-dependence, predicted by Halperin-Nelson equation, and can be expressed as $\xi=(2/A)\sinh(b/\sqrt{B_r})$, where $A$ and $b$ are parameters which are determined by fitting with experimental data.  By fitting the experimental data at $T=20$~mK and $V_g=170$~V, we obtain $A=1.8$, $b=0.2$, $C_1=1000$ and $C_2=1.65$. The data have been plotted in Fig. \ref{fig:Res_map}(e). Spatial inhomogeneity on the two-dimensional superconductor broadens the BKT transition \cite{PhysRevB.80.214506, daptary2016correlated} and a percolation transition is well accessible within the Halperin-Nelson theory.

The resistance at position ($x_i,y_i$) at a magnetic field $B$ and time $t$ is given by $R_i(B,t)=R_i(B)+\delta R_i(B,t)$. We start at $B=0.2$~T with $\delta R_i(B,t=0)=0$ and continuously update $R_i(B,t)$ at the interval of a relaxation time $\tau$ and finally reach the maximum field $B=2$~T. The amplitude of noise $\delta R_i(B,t)$ is chosen randomly from a set $\{ \delta R_i(B,t) \}$ of numbers which follows Gaussian distribution and has a standard deviation $0.001$ and zero mean. 
The statistics of the noise is, however, governed by the distribution of $\tau$ which is also chosen randomly from a set $\{ \tau_n \}$. We assume that the Josephson junctions, formed during the percolative superconducting transition, contribute non-Gaussian component in the resistance noise. We, therefore, consider that the distribution of the relaxation time has two components which can be expressed as $\{ \tau_n \}=x\{ \tau_n \}_{NGC}+(1-x)\{ \tau_n \}_{GC}$, $NGC$ stands for non-Gaussian component and $GC$ for Gaussian component. The fraction $x$, which defines the amount of non-Gaussianity in the noise, is taken to be proportional to the ratio of the superconducting region to the non-superconducting region. The distribution functions for $\{ \tau_n \}_{NGC}$ and $\{\tau_n \}_{GC}$ are determined by comparing the frequency-dependence of power spectral density (PSD) of resistance noise, given by the following equation, with the experimentally-obtained PSD:
\begin{equation}
	S_R(f)=\lim_{t_0\to\infty} \Big( \frac{1}{2t_0} \Big) \Big( \int_{-t_0}^{t_0} \delta R(t)e^{i2\pi ft}dt \Big)^2
\end{equation}
To incorporate the $1/f$-dependence of PSD and the influence of SOI, we include the second critical field $B_{c2||}$ in the PSD, through the following relation:
\begin{equation}
	S_R(f)=\int_{0}^{\infty} d\tau F(\tau) \frac{2\tau (B-B_{c2||})^3}{1+2\pi f\tau},
\end{equation}
where $F(\tau)$ is the distribution function for $\tau$. For the GC, we have a Gaussian distribution $F(\tau)=1/(\sqrt{2\pi \sigma^2})e^{-(\tau-\tau_{GC})^2/2\sigma^2}$, where $\sigma$ and $\tau_{GC}$ are, respectively, the variance and mean value of the Gaussian distribution. For the NGC, we use  a stretched exponential function $F(\tau) = 1/(2\sqrt{\pi}) \sqrt{\tau} e^{-\tau/\tau_{NGC}}$, typically used to study glassy dynamics. With $\tau_{GC}=\tau_{NGC}=500$~ns and $\sigma=100$~ns, the PSDs are calculated at different fields and the corresponding distributions of $\{ \tau_n \}$ are shown in Fig. \ref{fig:tau_dist}.

\begin{figure}[t!]
	\begin{center}
		\includegraphics[width=0.5\textwidth]{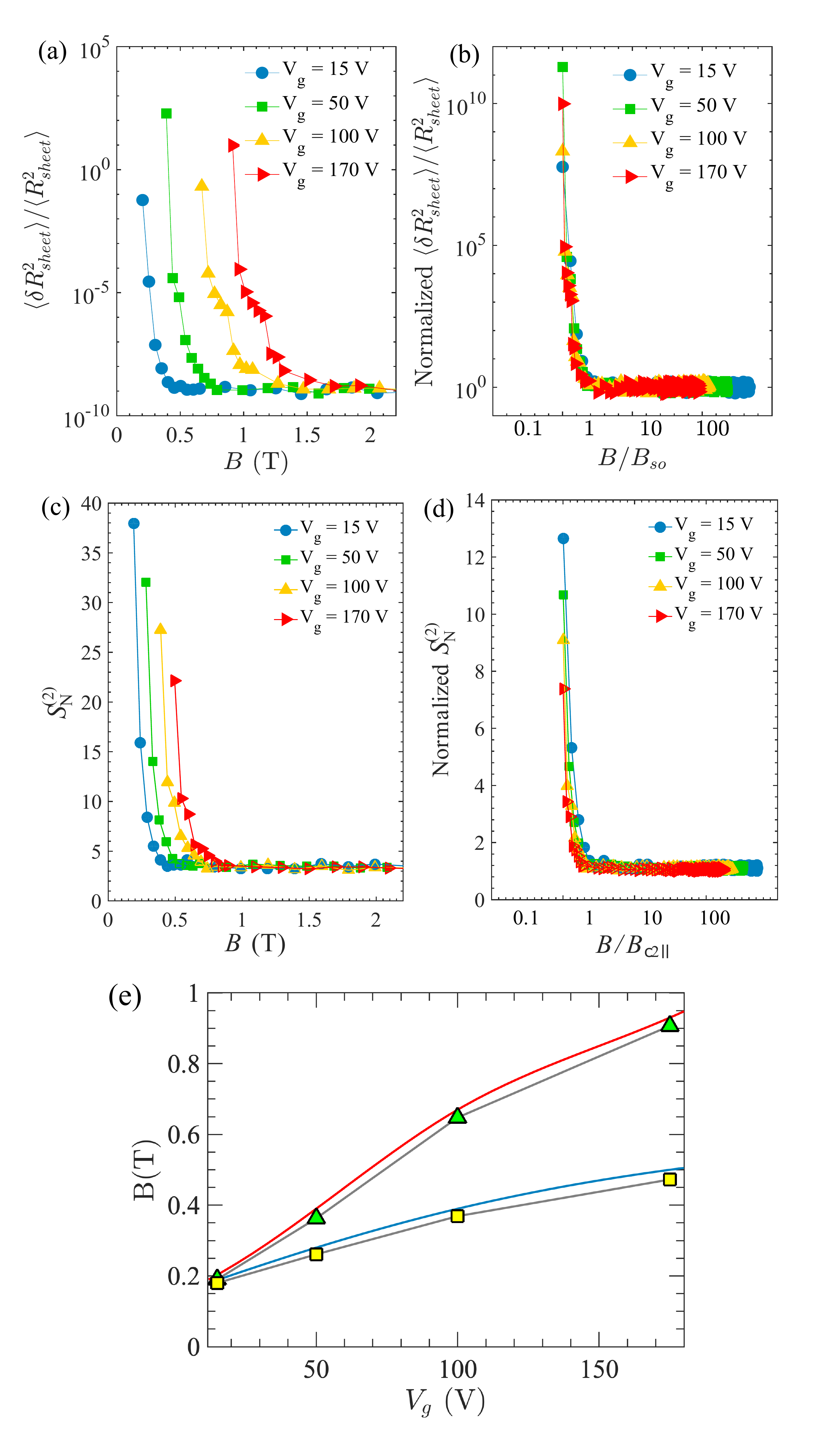}
		\small{\caption{ Gate-voltage variation of (a)  $\mathcal{R}$ with magnetic field $B$, (b) normalized  $\mathcal{R}$ with $B/B_{so}$, (c) $S_N^{(2)}$ with magnetic field $B$ and (d) normalized $S_N^{(2)}$ with $B/B_{c2||}$ at different values of gate-voltage $V_g$. The normalization of the quantities, plotted on the vertical axis, in (b) and (d) is performed using the respective values at the maximum value of the $B$ field. In this plot, temperature $T=20$~mK. (e) Variation of the critical fields $B_{c2||}$ (yellow squres) and $B_{so}$ (blue traingles) with gate-voltage. The modeled gate-voltage dependence is obtained from Fig.\ref{fig:figure6}. \label{fig:Noise_B}}}
	\end{center}
\end{figure}
The relative variance of the resistance fluctuations $\mathcal{R} \equiv \frac{\langle \delta R_{sheet}^2 \rangle}{\langle R_{sheet}^2 \rangle}$ and the normalized second spectrum $S_{N}^{(2)}$ are calculated by using Eq. \ref{PSD} and Eq. \ref{NSP} respectively. A plot of $\mathcal{R}$ and $S_{N}^{(2)}$ as a function of the field $B$ for different representative values of gate voltage $V_g$ are shown in Fig. \ref{fig:Noise_B}(a) and (c). The same set of obtained data, when plotted with respect to the field values, scaled using the critical fields $B_{so}$ and $B_{c2||}$, reveals that  $\mathcal{R}$ scales with $B_{so}$ while $S_{N}^{(2)}$ scales with $B_{c2||}$, as shown in Fig. \ref{fig:Noise_B}(b) and (d). The critical fields $B_{c2||}$ (yellow squres) and spin-orbit fields $B_{so}$ (blue traingles), obtained from the simulation are shown as a function of $V_g$ along with experiment  in Fig. \ref{fig:Noise_B}(e). The excellent match between experimental and simulation data tells that a simple random resistor network model is able to capture the essential features of resistance fluctuations close to the upper critical field in 2D inversion symmetry broken superconductors.\\
\section{CONCLUSION}
To conclude, we have probed, through careful measurements of  resistance fluctuations, the interplay of SOI, pairing potential  and Zeeman energy in the superconducting phase of LaAlO$_3$/SrTiO$_3$. We find the presence of larger non-Gaussian fluctuations below $B_{c2\parallel}$ arising due to correlated vortex-dynamics. Large, Gaussian resistance fluctuations were seen in the field range between  $B_{c2\parallel}$  and $B_{SO}$ which indicate the presence of superconducting clusters  without global phase coherence. We identify and quantify the relevant energy scales in this system -  SOI, Zeeman energy and pairing potential. Our work emphasizes the important role played by the interplay between  these energy scales in   framing the phase diagram of 2-D inversion asymmetric superconductors.
\section*{ACKNOWLEDGMENTS}
The authors thank R C Budhani, IIT Kanpur for 	providing the samples. AB acknowledges funding from SERB, DST, Govt. of India.  HKK acknowledges funding from CSIR, Govt. of India. 
\section*{APPENDIX}
\begin{figure}[t!]
\begin{center}
\includegraphics[width=0.42\textwidth]{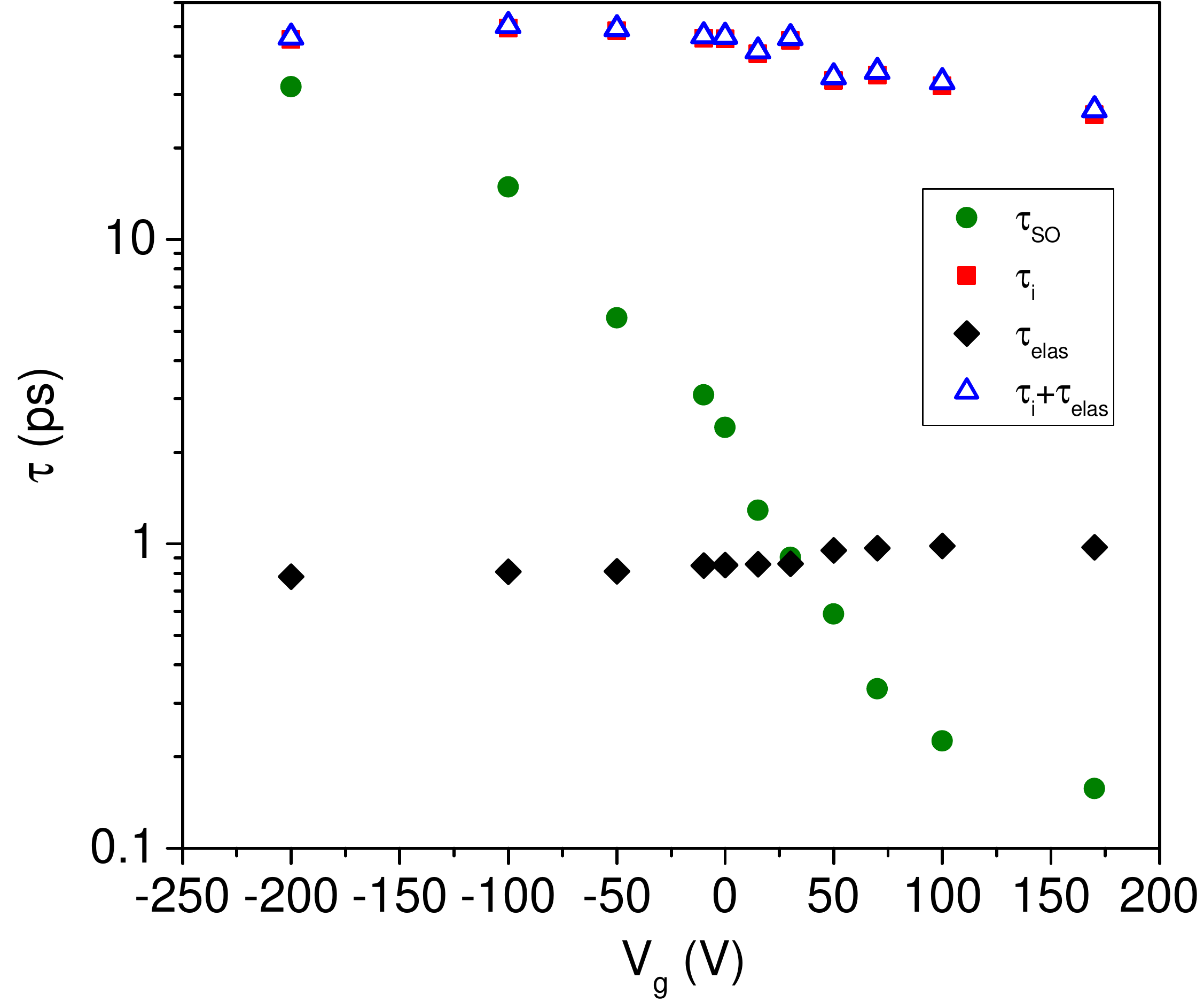}
\small{\caption{ Plot of spin-orbit scattering time $\tau_{SO}$ (olive filled circles), inelastic time $\tau_{i}$ (red filled squares), elastic time $\tau_{elas}$ (black filled) and total scattering time $\tau=\tau_{i}+\tau_{elas}$ (blue open triangles) versus $V_g$. \label{fig:appendix1}}}
\end{center}
\end{figure}

In Fig. \ref{fig:appendix1}, we plot the different scattering times extracted from Eq. \ref{Eqn:WAL} as a function of $V_g$ at $T=245$ mK. It can be seen that for all $V_g$, total scattering time $\tau$ (=$\tau_i$+$\tau_{elas}$, where $\tau_i$ and $\tau_{elas}$ are inelastic and elastic scattering time respectively) is larger than spin-orbit scattering time $\tau_{SO}$ implying strong spin-orbit interaction in the LaAlO$_3$/SrTiO$_3$ interface which are gate voltage tunable.



\end{document}